\documentclass[twocolumn,english,floatfix,superscriptaddress,prl,showpacs]{revtex4}
\usepackage[T1]{fontenc}
\usepackage[latin1]{inputenc}
\usepackage{amsmath}
\usepackage{graphicx}

\makeatletter
\@ifundefined{textcolor}{}
{%
 \definecolor{BLACK}{gray}{0}
 \definecolor{WHITE}{gray}{1}
 \definecolor{RED}{rgb}{1,0,0}
 \definecolor{GREEN}{rgb}{0,1,0}
 \definecolor{BLUE}{rgb}{0,0,1}
 \definecolor{CYAN}{cmyk}{1,0,0,0}
 \definecolor{MAGENTA}{cmyk}{0,1,0,0}
 \definecolor{YELLOW}{cmyk}{0,0,1,0}
 }

\usepackage{amsfonts}

\usepackage{amsfonts}

\makeatletter
\makeatother

\makeatother

\usepackage{babel}

\begin{document}

\title{Quantum ripples in strongly correlated metals}

\author{E. C. Andrade}

\affiliation{Instituto de Física Gleb Wataghin, Unicamp, C.P. 6165, Campinas,
SP 13083-970, Brazil}

\affiliation{Department of Physics and National High Magnetic Field Laboratory,
Florida State University, Tallahassee, FL 32306}

\author{E. Miranda}

\affiliation{Instituto de Física Gleb Wataghin, Unicamp, C.P. 6165, Campinas,
SP 13083-970, Brazil}

\author{V. Dobrosavljevi\'{c}}

\affiliation{Department of Physics and National High Magnetic Field Laboratory,
Florida State University, Tallahassee, FL 32306}

\date{\today}
\begin{abstract}
We study how well-known effects of the long-ranged Friedel oscillations
are affected by strong electronic correlations. We first show that
their range and amplitude are significantly suppressed in strongly
renormalized Fermi liquids. We then investigate the interplay of elastic
and inelastic scattering in the presence of these oscillations. In
the singular case of two-dimensional systems, we show how the anomalous
ballistic scattering rate is confined to a very restricted temperature
range even for moderate correlations. In general, our analytical results
indicate that a prominent role of Friedel oscillations is relegated
to weakly interacting systems.
\end{abstract}


\maketitle
\emph{Introduction.}---Fermi liquid theory is known to successfully
describe the leading low temperature behavior of metals, even in instances
of very strong correlations (e.g. heavy fermions \cite{stewart_nfl_rmp2001}).
In the presence of perturbations that break translational symmetry,
such as impurities and defects, the Fermi liquid re-adjusts itself
producing a spatially inhomogeneous pseudo-potential ``seen'' by quasiparticles
\cite{lee_ramakrishnan,NFL_2005}. Here the wave nature of the electrons
is manifested by the formation of ``ripples'', the Friedel oscillations
\cite{Friedel,crommie_nature_friedel_2d}, surrounding the perturbation.
Scattering processes of quasiparticles off these ripples then produce
new corrections to the $T$-dependence in transport quantities \cite{zna_prb01_linearT}. 

How significant are these corrections? The answer, of course, depends
on how broad the dynamic range is, in which such leading order non-analyticities
dominate. This question, as usual, cannot be answered by Fermi liquid
theory itself. What is needed is a microscopic model calculation that
is not restricted to obtaining the form of leading terms. A careful
and precise model calculation with such a goal is the central content
of this paper. 

We focus on single non-magnetic impurity scattering in an otherwise
uniform strongly interacting paramagnetic metal, where the analysis
is most straightforward and transparent, but this general issue is
of key relevance also for the diffusive regime. Our mostly analytical
results demonstrate that: (i) for sufficiently weak correlations we
recover the results of the Hartree-Fock approximation, in which the
effective scattering potential generated by the impurity is set by
the long-ranged Friedel oscillations; (ii) as we approach the Mott
transition, however, these oscillations are strongly suppressed as
the charge screening becomes more and more local, corresponding to
a shorter ``healing length''; (iii) a combination of ``healing'' and
inelastic scattering strongly suppresses the Friedel oscillation effects
even for moderate correlations. 

\begin{figure}[t]
\begin{centering}
\includegraphics[scale=0.3]{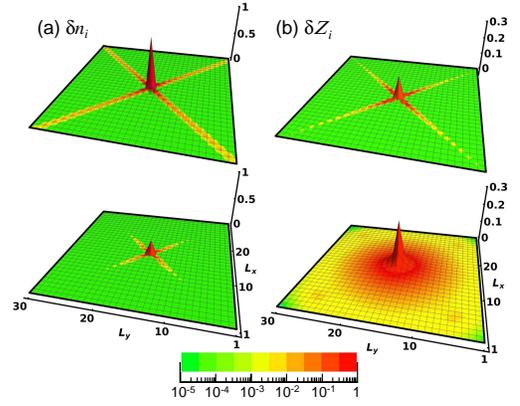} 
\par\end{centering}

\caption{\label{fig:delta_n_and_Z}\emph{(a)} Electronic density deviations
$\delta n_{i}=n_{i}-1$ (see text) displaying characteristic Friedel
oscillations. From top to bottom we have $m/m^{\star}=1.00$ and $0.30$.
The Friedel oscillations appear here as crosses because of the underlying
Fermi surface anisotropy \cite{Fridel_Fermi_Surface_Science09}. As
we enter the strongly correlated regime, these oscillations are suppressed,
similar to the ``healing'' effect found in Ref. \cite{nandini_healing}.
\emph{(b) }Quasiparticle weight deviations $\delta Z_{i}=Z_{i}-Z_{0}$.
From top to bottom we have $m/m^{\star}=0.85$ and $0.08$. While
for moderate values of interactions $\delta Z_{i}$ also displays
Friedel oscillations, it has a leading exponential decay close to
the Mott transition. Here, we have used a $30X30$ square lattice
with periodic boundary conditions and $\varepsilon_{o}=-D$. The color
scales encode only the positive values of both $\delta n_{i}$ and
$\delta Z_{i}$. }

\end{figure}

\emph{Model and method.---}We study the paramagnetic phase of the
disordered Hubbard model on a cubic lattice in $d$ dimensions \begin{eqnarray}
H & = & -\sum_{\left\langle ij\right\rangle ,\sigma}t_{ij}c_{i\sigma}^{\dagger}c_{j\sigma}+\sum_{i,\sigma}\varepsilon_{i}n_{i\sigma}+U\sum_{i}n_{i\uparrow}n_{i\downarrow},\label{eq:Hubbard_disordered}\end{eqnarray}
where $t_{ij}$ are the hopping matrix elements between nearest-neighbor
sites, $c_{i\sigma}^{\dagger}$$\left(c_{i\sigma}\right)$ is the
creation (annihilation) operator of an electron with spin projection
$\sigma$ at site $i$, $U$ is the on-site Hubbard repulsion, $n_{i\sigma}=c_{i\sigma}^{\dagger}c_{i\sigma}$
is the number operator, and $\varepsilon_{i}$ are the site energies.
All energies will be expressed in units of the clean half-bandwidth
(Fermi energy) $D$ and we approach the Mott metal-insulator transition
(MIT) by increasing $U$ at half filling (chemical potential $\mu=U/2$).
To treat this model, we employ the slave boson mean-field theory of
Kotliar and Ruckenstein \cite{kotliar_ruckenstein}, which is equivalent
to the Gutzwiller variational approximation \cite{brinkman_rice}.
In this approach, the renormalized site energies $v_{i}$ and the
local quasiparticle weights $Z_{i}$ are variationally calculated
through the saddle-point solution of the corresponding Kotliar-Ruckenstein
slave boson functional \cite{EGP2d_09}. This theory is mathematically
equivalent to a generalization of the dynamical mean field theory
(DMFT) \cite{dmft_rmp96} to finite dimensions, the statistical DMFT
(statDMFT) \cite{statdmft} implemented using a slave boson impurity
solver \cite{proceeding_sces08}. 

At $T=0$ and in the uniform limit $\left(\varepsilon_{i}=0\right)$
we have $v_{0}=0$ and $Z_{0}=1-u^{2}$ , with $u=U/U_{c}$ \cite{kotliar_ruckenstein}.
The critical interaction value $U_{c}$ for which the Mott metal-insulator
transition occurs is characterized by the divergence of the effective
mass $m^{\star}=m/Z_{0}$, where $m$ is the electron band mass, indicating
the transmutation of all electrons into localized magnetic moments.

We consider first a generic weak disorder potential $\left(\left|\varepsilon_{i}\right|\ll D\right)$
and expand the resulting mean-field equations, Eqs. $\left(5\right)$
and $\left(6\right)$ from \cite{proceeding_sces08}, around the uniform
solution. For particle-hole symmetry $Z_{i}=Z_{0}+\emph{O}\left(\varepsilon_{i}^{2}\right)$
and, up to first order in $\varepsilon_{i}$, the renormalized disorder
potential, which is the effective potential ``seen'' by the quasiparticles
at the Fermi level, reads (summation over repeated indices implied
throughout)\begin{eqnarray}
v_{i} & = & \left[1-u^{2}\right]^{-1}\left(\varepsilon_{i}-\left[\mathbf{M}^{-1}\left(u\right)\right]_{ij}\varepsilon_{j}\right).\label{eq:ren_site_energy_gen}\end{eqnarray}
The matrix $\mathbf{M}\left(u\right)$ is the lattice Fourier transform
of\begin{eqnarray}
M_{\mathbf{q}}\left(u\right) & = & 1-2g\left(u\right)\left[U_{c}\Pi_{\mathbf{q}}^{\left(0\right)}\right]^{-1},\label{eq:M_matrix_def}\end{eqnarray}
where $\Pi_{\mathbf{q}}^{\left(0\right)}$ is the usual static Lindhard
polarization function \cite{Mahan_many_body}, of the clean, non-interacting
system and\begin{eqnarray}
g\left(u\right) & = & \left(1+u\right)\left(1-u\right)^{2}\left[2u+u^{2}\left(1-u\right)\right]^{-1}.\label{eq:g_func_def}\end{eqnarray}

Charges rearrange themselves to screen the impurity potential and
the local electronic density is given by $n_{i}=1+\delta n_{i}$,
where\begin{eqnarray}
\delta n_{i} & = & -4g\left(u\right)\left[U_{c}\left(1-u^{2}\right)\right]^{-1}\left[\mathbf{M}^{-1}\left(u\right)\right]_{ij}\varepsilon_{j},\label{eq:delta_n_gen}\end{eqnarray}
with the same spatial structure as $v_{i}$ in Eq. \eqref{eq:ren_site_energy_gen}. 

We particularize now to a single impurity with energy $\varepsilon_{o}$
placed at the site $o$ such that $\varepsilon_{i}=\delta_{i,o}\varepsilon_{o}$.
Although obtained for weak scattering, our analytical theory does
capture the qualitative trends even when the scattering is not weak,
as we show numerically, see Fig. \eqref{fig:delta_n_and_Z}.

\emph{Weak and strong coupling limits}.--- In the weak coupling regime
$\left(u\rightarrow0\right)$ there is no mass renormalization and
Eqs. \eqref{eq:ren_site_energy_gen} and \eqref{eq:delta_n_gen} agree
with the Hartree-Fock solution of the Hubbard model, with a local
(static) self-energy given by $\Sigma_{i}=Un_{i}^{0}$, where $n_{i}^{0}=1+2\Pi_{ij}^{\left(0\right)}\varepsilon_{j}$
is the non-interacting electronic density and $\Pi_{ij}^{\left(0\right)}$
is the lattice Fourier transform of $\Pi_{\mathbf{q}}^{\left(0\right)}$.
Even though the bare impurity potential $\varepsilon_{i}$ is localized
in space, the density deviation $\delta n_{i}^{0}$ displays long-ranged
Friedel oscillations encoded in $\Pi_{ij}^{\left(0\right)}$. For
example, in a free electron gas we have $\delta n\left(r\right)\sim\mbox{cos}\left(2k_{F}r\right)/r^{d}$,
where $r$ is the distance to the impurity and $k_{F}$ is the Fermi
momentum. Slowly decaying Friedel oscillations are a direct consequence
of the gapless nature of Fermi liquid excitations. The renormalized
disorder potential reads $v_{i}\simeq\varepsilon_{i}+U\Pi_{ij}^{\left(0\right)}\varepsilon_{j}$,
implying that the electrons scatter not only off the local bare impurity,
but also off the long-ranged potential generated by the Friedel oscillations. 

As we approach the critical region $\left(u\rightarrow1\right)$,
however, the density deviation in Eq. \eqref{eq:delta_n_gen} becomes
\begin{eqnarray}
\delta n_{i} & = & -\frac{2\left(1-u\right)}{U_{c}}\left[\varepsilon_{i}+\frac{2}{U_{c}}\left(1-u\right)^{2}\left[\mbox{\boldmath\ensuremath{\Pi^{\left(0\right)}}}\right]_{ij}^{-1}\varepsilon_{j}\right],\label{eq:delta_n_critical}\end{eqnarray}
showing a suppression of the Friedel oscillations: the non-local part
of $\delta n_{i}$ is a factor $\left(1-u\right)^{2}$ smaller than
the local one. Therefore, the electronic density is significantly
disturbed only in the vicinity of the impurities, implying a much
shorter ``healing length'', see Fig. \eqref{fig:delta_n_and_Z}(a).
The suppression of the slow spatial decay in $\delta n_{i}$ reflects
the fundamental tendency of quasiparticles to become localized as
the system approaches the Mott insulator. 

The renormalized disorder potential, Eq. \eqref{eq:ren_site_energy_gen},
is \begin{eqnarray}
v_{i} & = & -\left(1-u\right)U_{c}^{-1}\left[\mbox{\boldmath\ensuremath{\Pi^{\left(0\right)}}}\right]_{ij}^{-1}\varepsilon_{j},\label{eq:v_critical}\end{eqnarray}
and the screened impurity potential is just as non-local as for small
$u$, except for a reduction of the overall amplitude scale. Therefore,
we should not be guided by density fluctuations alone which are indeed
healed very effectively. However, we notice that $v_{i}$ goes to
zero linearly at all lattice sites at the transition, signaling a
complete suppression of disorder by interactions \cite{screening_2003}.

To obtain the leading energy correction of $Z_{i}$, we have to expand
the mean-field equations up to second order in $\varepsilon_{i}$.
At intermediate values of the interaction, deviations in the quasiparticle
weights $\delta Z_{i}=Z_{i}-Z_{0}$ also show Friedel-like oscillations.
Close to the Mott transition, $\delta Z_{i}$ displays a leading exponential
decay from the impurity site, since all additional terms describing
its long-range oscillations are of higher order in $\left(1-u\right)$,
see Fig. \eqref{fig:delta_n_and_Z}(b) and Eq. \eqref{eq:delta_Z_critical}
below (the details of the calculation will be presented elsewhere).
A finite impurity potential tends to push the site occupation away
from half filling, thus reducing the tendency to form a local moment
and rendering the given site locally more metallic by increasing $Z_{i}$.
As spatial correlations grow, this ``metallization'' of the correlated
metal tends to spread out away from the impurity, thus creating metallic
``puddles'' (red region in Fig. \eqref{fig:delta_n_and_Z}(b)) in
an almost-localized host. A somewhat similar result emerges in the
$t-$$t^{\prime}-$$J$ model, in which an impurity induces a local
staggered magnetization in its vicinity \cite{Hirschfeld_SB2_nonmag_impurities,review_hirschfeld}
whose spatial extent also increases with correlations. The critical
behavior of $\delta Z_{i}$ is captured by our analytical expressions
and we can show that, for $d\ge2$, and $r_{io}/\xi\gg1$ \begin{eqnarray}
\delta Z_{i} & \sim & \frac{1-u}{U_{c}^{2}}\left(\frac{\pi^{\left(1-d\right)/2}}{2^{\left(1+d\right)/2}\xi^{\left(d-3\right)/2}}e^{-r_{io}/\xi}\right.\nonumber \\
 & - & \left.4\left(1-u\right)^{3}\left[\mbox{\boldmath\ensuremath{\Pi^{\left(0\right)}}}\right]_{io}^{-1}\right)\varepsilon_{o}^{2}\label{eq:delta_Z_critical}\end{eqnarray}
where $r_{io}=\left|\mathbf{r}_{i}-\mathbf{r}_{o}\right|$, $z$ is
the coordination number, and $\xi=\left(2z(1-u)\right)^{-1/2}$ plays
the role of a correlation length. This correlation length diverges
at the transition with a mean field exponent $1/2$. Previous studies
on the interface of a strongly correlated metal and a Mott insulator
\cite{Rosch_Mott_insulator,Tosatti_dead_layer}, which use techniques
similar to ours, also find an analogous leading exponential decay
of the quasiparticle weight upon entering the Mott insulator from
the metal. In those studies, however, the oscillating terms of Eq.
\eqref{eq:delta_Z_critical} seem to have been overlooked. 

\emph{Leading finite T corrections for transport and inelastic cutoffs.}---We
would like now to go beyond $T=0$ and study the behavior of the leading
temperature corrections to the resistivity $\rho\left(T\right)$ as
a function of the correlations. We focus henceforth only on $2d$
systems, since, in the weakly correlated regime, electron scattering
by Friedel oscillations leads to a non-Fermi-liquid linear temperature
correction to $\rho\left(T\right)$ \cite{zna_prb01_linearT,Review_Kravchenko_Sarachik04}
in the ballistic regime.

The transport scattering rate is given by\begin{eqnarray}
\tau_{\mbox{tr}}^{-1}\left(\varepsilon\right) & = & n_{imp}m\int_{0}^{2\pi}\frac{d\theta}{2\pi}\left|T_{q}\right|^{2}\left(1-\mbox{cos}\theta\right),\label{eq:transposrt_rate_def}\end{eqnarray}
where $n_{imp}$ is the impurity concentration, $T_{q}$ is the Fourier
transform of the $T$ matrix, $q=2k\mbox{sin}\theta/2$ is the transferred
momentum, $\theta$ is the scattering angle. To simplify our analytical
expressions, we henceforth focus on the case a free electron dispersion
$\varepsilon=k^{2}/2m$; we carefully checked that no significant
changes are found if a different dispersion is used. Up to first order
in the impurity potential $\varepsilon_{o}$, the $T$ matrix is simply
given by the renormalized disorder potential $v_{i}$ (Eq. \eqref{eq:ren_site_energy_gen}).
The scattering time is given by the average $\tau_{\mbox{tr}}=\int d\varepsilon\tau_{\mbox{tr}}\left(\varepsilon\right)f^{\prime}\left(\varepsilon\right)$,
where $f^{\prime}\left(\varepsilon\right)$ is the derivative of the
Fermi distribution function.

In our slave boson mean-field theory, the electronic self-energy is
purely real \cite{kotliar_ruckenstein} and describes only the elastic
scattering of the electrons off a temperature dependent screened impurity
potential. However, this scheme should really be regarded as a variational
calculation of the quasiparticle parameters within our statDMFT procedure.
In a fuller treatment, there is also an imaginary part in the self-energy,
reflecting inelastic effects. For the purposes of examining the \emph{leading
perturbative effects} of impurity scattering, the imaginary part can
be computed in the uniform system, where it emerges naturally in the
context of local Fermi liquid theories like DMFT \cite{dmft_rmp96,Kadowaki_Woods_general_NatPhys09}
and is given by\begin{eqnarray}
\gamma\left(T\right) & = & \Lambda\left(u\right)T_{F}\left(T/T_{F}\right)^{2},\label{eq:gamma_T_def}\end{eqnarray}
where $T_{F}$ is the Fermi temperature and the function $\Lambda\left(u\right)$
has the following limits: $\Lambda\left(u\right)\sim u^{2}$ for small
$u$ and $\Lambda\left(u\right)\sim\left(1-u\right)^{-2}\sim\left(m^{\star}/m\right)^{2}$
close to the Mott transition. These limits can be understood from
the fact that in the weakly correlated regime inelastic scattering
effects are perturbative, whereas in the strongly correlated regime
we recover the well known Kadowaki-Woods relation \cite{Kadowaki_Woods_general_NatPhys09},
observed in several strongly correlated systems, and which holds within
the DMFT picture we use. 

There are two leading contributions from inelastic scattering. A bulk
one, present even in the clean limit, given by $\tau_{in}^{-1}\left(T\right)=\eta\gamma\left(T\right)$,
where $\eta$ is a geometrical factor depending on the band structure
used in the DMFT calculation. $\tau_{in}^{-1}\left(T\right)$ simply
adds to $\tau_{\mbox{tr}}^{-1}\left(T\right)$ in Eq. \eqref{eq:transposrt_rate_def}
through Matthiessen's rule, since we consider very dilute impurities.
In addition to this, a finite imaginary part also \emph{cuts off the
leading non-analyticities of the elastic scattering off Friedel oscillations}
\cite{zna_prb01_linearT,das_sarma_lindhard2d}. This is taken into
account in the calculation of $\Pi_{\mathbf{q}}^{\left(0\right)}$,
considering that the electron energy has now an imaginary part given
by $\gamma\left(T\right)$. The calculation of $\Pi_{\mathbf{q}}^{\left(0\right)}$
in the presence of inelastic broadening is carefully discussed in
Ref. \cite{das_sarma_lindhard2d} for the $2d$ electron gas and we
use the analytical form of $\Pi_{\mathbf{q}}^{\left(0\right)}$ as
obtained there. 

The final form of $\tau_{\mbox{tr}}^{-1}\left(T\right)$, valid for
$T\ll T_{F}$, reads\begin{eqnarray}
\tau_{\mbox{tr}}^{-1}\left(T\right) & = & \tau_{0}^{-1}A^{2}\left(u\right)\left\{ 1+2\frac{T}{T_{F}}\alpha\left(u\right)w\left(T,\gamma\left(T\right)\right)\right\} \nonumber \\
 & + & \eta\gamma\left(T\right),\label{eq:scattering_rate_final}\end{eqnarray}
where\begin{eqnarray}
w\left(T,\gamma\right) & = & \int_{-\infty}^{+\infty}\frac{dx}{4}\mbox{sech}^{2}\left(\frac{x}{2}\right)\mbox{Re}\left[\mbox{ln}\Gamma\left(\frac{2\pi}{\pi+\frac{\gamma\left(T\right)}{T}+ix}\right)\right]\nonumber \\
 & + & \frac{1}{2}\mbox{ln}\left(2\pi\right)+\frac{\gamma\left(T\right)}{2\pi T}\mbox{ln}\left(\frac{T_{F}}{2\pi T}\right),\label{eq:w_def}\end{eqnarray}
\begin{eqnarray}
A\left(u\right) & = & g\left(u\right)\left[\left(1-u^{2}\right)\left(\rho\left(\varepsilon_{F}\right)U_{c}+g\left(u\right)\right)\right]^{-1},\label{eq:A_func_def}\end{eqnarray}

\begin{eqnarray}
\alpha\left(u\right) & = & 2\rho\left(\varepsilon_{F}\right)U_{c}\left[\rho\left(\varepsilon_{F}\right)U_{c}+g\left(u\right)\right]^{-1},\label{eq:alpha_def}\end{eqnarray}
$\tau_{0}^{-1}$ is the zero-temperature impurity scattering rate,
$\rho\left(\varepsilon_{F}\right)$ is the clean electronic density
of states at the Fermi level and $\Gamma\left(z\right)$ is the Gamma
function. The function $A\left(u\right)$ controls the amplitude of
the scattering rate from the screened impurity potential. In the weakly
interacting regime $A\left(u\right)\sim1$, whereas in the critical
region $A\left(u\right)\sim\left(1-u\right)$ due to a vanishing $v_{i}$
at $U=U_{c}$, Eq. \eqref{eq:v_critical}. The function $\alpha\left(u\right)$
gives the strength of the leading temperature correction. It goes
as $U$ for weak correlations, indicating that the temperature corrections
only arise in the presence of electron-electron repulsion, and saturates
to $2$ close to the Mott transition. The function $w\left(T,\gamma\right)$
encodes the dependence of the leading temperature correction on $\gamma\left(T\right)$.

For $\gamma\left(T\right)=0$, only elastic scattering is present
and we obtain $w\left(T,0\right)=0.5$. Plugging this into Eq. \eqref{eq:scattering_rate_final},
we see that the linear in $T$ correction is found for all $u>0$
\cite{zna_prb01_linearT} and is limited only by the overall amplitude
$A\left(u\right)$, which is in accordance with Eq. \eqref{eq:v_critical}. 

\begin{figure}[t]
\begin{centering}
\includegraphics[bb=53bp 40bp 714bp 530bp,scale=0.3]{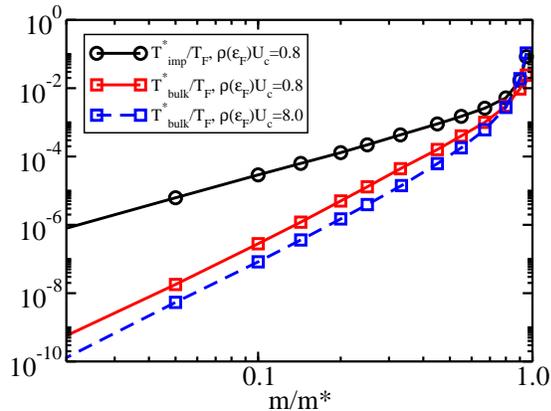} 
\par\end{centering}

\caption{\label{fig:Limiting-temperature}Limiting temperatures to the linear
in $T$ regime of the resistivity of a $2d$ electron liquid on a
log-log scale, calculated under two different assumptions (see text)
for two values of the parameter $\rho\left(\varepsilon_{F}\right)U_{c}$
\cite{rho(E_F)Uc}. $T_{imp}^{\star}$ is almost independent of $\rho\left(\varepsilon_{F}\right)U_{c}$
and only one curve is shown here. $T_{bulk}^{\star}$ is the dominating
cutoff above which the linear behavior is lost. Here, we have used
$\tau_{0}\varepsilon_{F}=10$, corresponding to a weak impurity potential. }

\end{figure}

However, for a finite $\gamma\left(T\right)$, the linear region of
$\rho\left(T\right)$ considerably narrows as we enter the correlated
regime. As there are two leading contributions from inelastic scattering,
we analyze their individual effects separately by defining two threshold
temperatures bounding the non-Fermi-liquid region from above. They
are obtained by comparing the purely elastic $\tau_{\mbox{tr}}^{-1}\left(T\right)$
($\gamma\left(T\right)\to0$ in Eq. \eqref{eq:scattering_rate_final})
with two other scattering rates: one in which we let $\eta\to0$ ($T_{imp}^{\star}$)
and the other in which we set $w\left(T,\gamma\right)\to w\left(T,0\right)$
($T_{bulk}^{\star}$) in Eq. \eqref{eq:scattering_rate_final}. We
interpolate the function $\Lambda\left(u\right)$ using DMFT with
both quantum Monte Carlo and iterated perturbation theory \cite{Merino_McKenzie_DMFT_transport}
as impurity solvers. We use $\eta=10$, a value also in agreement
with DMFT calculations \cite{Merino_McKenzie_DMFT_transport}. As
we can see from Fig. \eqref{fig:Limiting-temperature}, $T_{bulk}^{\star}$
is strictly smaller than $T_{imp}^{\star}$. This is because $T_{bulk}^{\star}$
is not only proportional to the infinitesimal impurity concentration
$n_{imp}$ (or, equivalently, to $\left(\tau_{0}\varepsilon_{F}\right)^{-1}$)
but, in the critical region, $T_{bulk}^{\star}\sim\left(m/m^{\star}\right)^{4}$,
whereas $T_{imp}^{\star}\sim\left(m/m^{\star}\right)^{2}$. Thus,
for any degree of correlations, it is $T_{bulk}^{\star}$ which sets
the upper bound on the linear in $T$ region of $\rho\left(T\right)$.
From Fig. \eqref{fig:Limiting-temperature} we see that, even for
very moderate correlations, e.g. for $m/m^{\star}\sim0.9$, $T_{bulk}^{\star}\sim10^{-2}T_{F}$,
and for $m/m^{\star}\sim0.6$, already $T_{bulk}^{\star}\sim10^{-4}T_{F}$.
Thus, the non-Fermi liquid region is limited to very low temperatures.
Ultimately, as the linear in $T$ regime is also bounded from below
by a crossover to the diffusive regime, the ballistic $T$-interval
in which these elastic corrections dominate may not be present at
all.

\emph{Conclusions}.---We presented a detailed, mostly analytical,
model calculation of the effects of a single non-magnetic impurity
placed in a correlated host. We find that strong correlations tend
to reduce the effects of the long-range part of the Friedel oscillations
and our work provides clear analytical insight into how this happen.
It should be possible to directly test our quantitative predictions
by means of current generation scanning tunneling microscopy methods,
shedding new light on the behavior near the Mott MIT. It is noteworthy
that impurities placed in d-wave superconductors also produce slowly
decaying perturbations in real space, reflecting their gapless nature,
through a mechanism closely related to Friedel oscillations in normal
metals. Recent work by Garg \emph{et al.} \cite{nandini_healing}
shows, by using a very similar theoretical approach as we do, that
in this system strong correlations also lead to spatial ``healing''.
We believe that both phenomena have a closely related origin and our
results strongly suggest that the ``healing'' effect is a more general
property of correlated metals close to the Mott transition, not an
effect specific to cuprates or the superconducting state. 

This work was supported by FAPESP through grants 04/12098-6 (ECA)
and 07/57630-5 (EM), CAPES through grant 1455/07-9 (ECA), CNPq through
grant 305227/2007-6 (EM), and by NSF through grant DMR-0542026 (VD).


\begin{thebibliography}{25}
\bibitem[{rho()}]{rho(E_F)Uc}
\bibinfo{note}{Within our slave boson approach,
  $U_{c}=8\left|\varepsilon_{kin}^{0}\right|$, where $\varepsilon_{kin}^{0}$ is
  the average kinetic energy in the non-interacting limit. Therefore, for the
  band structures used here, we have
  $\rho\left(\varepsilon_{F}\right)U_{c}\sim\mathcal{O}\left(1\right)$.}
\end{thebibliography}

\end{document}